**Electron transfer in confined electromagnetic fields**

Alexander Semenov[1] and Abraham Nitzan[1,2] [*]

[1] Department of Chemistry, University of Pennsylvania, Philadelphia, Pennsylvania 19104, USA
[2] School of Chemistry, The Sackler Faculty of Science, Tel Aviv University, Tel Aviv 69978, Israel

**Abstract**

The interaction between molecular (atomic) electron(s) and the vacuum field of a reflective cavity generates a significant interest thanks to the rapid developments in nanophotonics. Such interaction which lies within the realm of cavity quantum electrodynamic can substantially affect transport properties of molecular systems. In this work we consider non-adiabatic electron transfer process in the presence of a cavity mode. We present a generalized framework for the interaction between a charged molecular system and a quantized electromagnetic field of a cavity and apply it to the problem of electron transfer between a donor and an acceptor placed in a confined vacuum electromagnetic field. The effective system Hamiltonian corresponds to a unified Rabi and spin-boson model which includes a self-dipole energy term. Two limiting cases are considered: one where the electron is assumed much faster than the cavity mode and another in which the electron tunneling time is significantly larger than the mode period. In both cases a significant rate enhancement can be produced by coupling to the cavity mode in the Marcus inverted region. The results of this work offer new possibilities for controlling electron transfer processes using visible and infrared plasmonics.

[*] Author whom all correspondence should be addressed; electronic mail: anitzan@sas.upenn.edu

# 1. Introduction and background

Controlling, manipulating and modifying transport properties on the nanoscale is key to creating useful nanodevices[1–3], and can be pursued in different ways by creating suitable structures, changing environmental interactions[4] (including the electromagnetic (EM) environment[5,6]) or imposing thermal[7] or electric potential gradients[8]. Of recent interest in several fields is the effect of specific coupling to a bosonic environment, regardless of its nature, on electronic (charge or energy) transport processes. The concept itself is of course not new: arguably the most prominent example is electron transfer (ET) in condensed molecular systems, where nuclear motion (including solvent reorganization) is needed to bring the system to a transition configuration in the donor/acceptor subspace on one hand, and to stabilize the reaction product(s) on the other[9]. In this and many other molecular processes where the thermal environment plays an important active role, the latter is modeled as a harmonic bath, yielding variants of the spin-boson model.

Electron transfer can be also affected by vibrational motions in the bridge connecting between donor and acceptor. Even weak effects of this kind may be directly observed by inelastic tunneling spectroscopy[10–12]. They become significant when the transmission strongly depends on the instantaneous nuclear configuration[13–16] or when dephasing by thermal nuclear motion causes transition from coherent to diffusive transport behavior.[17] Active control of electron transfer has recently been demonstrated by several experimental works, which show that IR excitation of bridge nuclear motions can significantly alter ET process in donor-bridge-acceptor system[18,19,20]. These observations are usually attributed to the modulation of the effective donor-acceptor (DA) coupling (electron tunneling probability) on the underlying nuclear configuration which is in turn affected by vibrational excitation. Such effects can be considerably enhanced if this modulation affects a process dominated by interference between transfer paths, e.g. when ET unassisted by IR radiation is forbidden because of destructive interference between ET pathways.[21,22] Such "destruction of destructive interference" has been implicated also in the observed temperature dependence of electric current[23].

Other generic mechanisms by which a bosonic environment can determine, or at least strongly affect, electronic transport phenomena are associated with some kind of resonance between electronic and vibrational energies. One such set of phenomena of strong current interest is the long-standing issue of the high efficiency of electronic energy transfer (exciton motion) in photosynthetic materials. While the precise mechanism of this process is still not fully established, many studies suggest that the coupling between chromophores and inter/intramolecular vibrations can play a major role in the observed coherences between electronic states[24] and in the observed efficient exciton transport in some systems[25]. A series of theoretical studies[26–31] supports this idea. A generic Hamiltonian used in such studies is

$$\hat{H} = \sum_{i=1}^{N} \{E_i + \sum_l \lambda_{i,l}(\hat{b}_l^\dagger + \hat{b}_l)\}\hat{c}_i^\dagger \hat{c}_i + \sum_{i,j=1}^{N} J_{ij}\hat{c}_i^\dagger \hat{c}_j + \sum_l \hbar\omega_l \hat{b}_l^\dagger \hat{b}_l \qquad (1)$$

where $\hat{c}_i^\dagger (\hat{c}_i)$ correspond to excitation creation(annihilation) operator for $i^{th}$ molecular site (chromophore), $\hat{b}_l^\dagger (\hat{b}_l)$ are creation(annihilation) for $l^{th}$ vibration, $E_i$ denotes the energy of $i^{th}$ chromophore, $\lambda_{i,l}$ is a vibronic coupling constant, $J_{ij}$ is a dipole coupling between chromophores. When the resonance condition

$$\hbar\omega_l \approx E_i - E_j \qquad (2)$$

is met, the mode $l$ and the chromophores $i$ and $j$ can be strongly coupled and form a mixed delocalized vibronic state which provides an effective pathway for energy exchange between chromophores. Here, vibrational energy is used to bridge the gap between electronic excitations otherwise localized on different molecular centers[32]. The contribution of such a pathway is particularly significant in situations where the usual *local* hopping is suppressed, e.g. due to disorder. In addition to bridging electronic energy gaps, the localized or delocalized nature of the relevant vibrational modes has been discussed[13,27,30] and it was pointed out that delocalized vibrations can be significant exciton coherence.

Besides molecular vibrations, the radiation field provides another kind of a bosonic environment that couples to and may significantly affect molecular processes. Traditional photochemistry may be enhanced by active coherent control, where the molecular time evolution is manipulated by tailored coupling to the radiation field whose high intensity often allows a classical treatment[33]. However, under strong light-matter coupling conditions that can be realized for molecules placed in optical cavities, even a vacuum radiation field can strongly affect energy and electron transport in the molecular system[34–36]. Studies of such phenomena, that lie within the realm of cavity quantum electrodynamics[37], have recently generated significant experimental and theoretical interest enhanced by the rapid developments in nanophotonics. The common underlying mechanism of these phenomena is the energy exchange between the photonic and molecular degrees of freedom through dipole coupling, leading to the formation of mixed (hybridized) photon-matter states – polaritons. This mixing is not only manifested in optical spectra[38,39] but can modify potential surfaces associated with excited electronic states with predicted consequences for photodissociation[40–42] and photoinduced electron transfer[43–45], as well as the dynamics of charge and energy transfer in molecular systems[45–51]

These phenomena stem from three elements that characterize molecular systems coupled to cavity modes: First, the coupling, $\mu_{eg}\sqrt{\omega/2\varepsilon_0\Omega}$, where $\mu_{eg}$ is the molecular transition dipole element and ω is the molecular transition frequency, is relatively strong in cavities of small volume Ω. Second, the molecular transition frequency ω is assumed to be in resonance with the fundamental cavity-photon. In principle the molecular frequency ω can be the energy needed to bridge the gap between electronic excitations on different molecular sites (a mechanism akin to the one discussed above of enhancing exciton transfer by bridging the energy gap between different molecular excitations, however most applications discussed to date consider cavity photons in resonance with the exciton itself, i.e., the energy difference between the ground and excited molecular states. Third, strong coupling of the cavity photon to an electronic transition leads to a local renormalization of the nuclear potential energy surfaces mainly near nuclear configurations for which the electronic transition energy is in resonance with the cavity photon. Fourth, the coupling of excitons or charge carrier to the cavity photon, which by its nature is delocalized in the cavity, can increase the coherence length of these energy and charge carriers and enhance their mobility in particular in situations where the mobility is otherwise reduced by disorder[37, 40, 44]. Similar enhancement of exciton transport by coupling to surface plasmons has been extensively discussed, see, e.g. Refs. 52-54 . Yet another important implication of the coupling of possibly many molecules to a single cavity photon mode is the possible appearance of collective effects where system properties depend on the number of molecules in a non-additive way. Such phenomena, well known in observations of superradiance[55] and superfluorescence[56], are often observed in the Rabbi splitting that characterizes avoided crossing phenomena associated with strong exciton-plasmon/cavity mode coupling[38,39], were suggested to affect other dynamical aspects of molecular aggregates in cavity environments[57,58].

The Hamiltonians used for modeling exciton (or charge carrier)-cavity photon dynamics, e.g.[40],

$$\hat{H} = \sum_{i=1}^{N} E_i \hat{c}_i^\dagger \hat{c}_i + \sum_{i,j=1}^{N} t_{ij} \hat{c}_i^\dagger \hat{c}_j + \sum_{i=1}^{N} \Omega_i \times (\hat{a}^\dagger + \hat{a})(\hat{c}_i^\dagger + \hat{c}_i) + \hbar\omega \hat{a}^\dagger \hat{a} \qquad (3)$$

where $\hat{c}_i^\dagger$ $(\hat{c}_i)$ create (annihilate) excitation on site $i$, $t_{ij}$ is a dipole- dipole interaction between excitonic pairs, $\hbar\omega \hat{a}^\dagger \hat{a}$ is a cavity mode and $\Omega_i \times (\hat{a}^\dagger + \hat{a})(\hat{c}_i^\dagger + \hat{c}_i)$ is a dipole coupling describing energy exchange between the molecule and the cavity mode, are similar to those like the Hamiltonian (1) used to describe exciton-vibration coupling in light harvesting systems. Important differences between these systems should however be noted: First, the high frequencies of optical modes supported by nanocavities make them operate at effectively zero temperature, thus emphasizing quantum effects. Secondly, standard considerations of (approximately) harmonic motions interacting with electronic dynamics are usually associated with the mutual effects of interacting electronic and nuclear dynamics, where the focus is on the timescale separation between these motions and the manifestations of events where it breaks down. Consequently, the resonances promoting exciton transfer are between vibrational frequencies and energy *differences* between different molecular excitons. In contrast, the harmonic optical modes are often tailored to be in resonance with electronic excitations, implying dynamics on similar timescales. We note in passing that recent studies of vibrational strong coupling indicate that molecular nuclear dynamics in the ground electronic states may be affected by coupling to the radiation field in cavities that support infrared photon modes[57,59–66]. Similar cavity modes can be used to affect exciton motion by bridging excitonic energy gaps as in Refs. 24-32.

The strong light-matter coupling associated with cavity confined modes implies that cavity environment may have observable, perhaps strong, effects on molecular properties even in the absence of incident radiation, that is, for molecules interacting with the vacuum cavity field. Common to the cavity phenomena described above is that they take place in states where enough energy is available to excite the cavity mode. In the present work we consider non-adiabatic electron transfer process between a donor and an acceptor in the presence of a cavity mode, where such energy is not necessarily available. The question addressed is, can such processes be affected by coupling to the cavity mode, and if so to what extent? The paper is organized as follows: in Section 2 we derive the model Hamiltonian and examine its implications for the electron transfer problem. In Section 3 the developed framework is used to analyze ET processes in model systems where coupling to the EM environment can potentially have a strong effect on the transfer rate and examine the effects with respect to variations of different system parameters. We conclude in Section 4 where potential applications of our results are discussed.

## 2. Theoretical Framework

Consider an electron (sometimes referred to as the excess electron) that interacts with a neutral system of charges (e.g., bound molecular electrons and nuclei), together referred to henceforth as a molecular system, placed inside an electromagnetic cavity of volume $\Omega$ and frequency $\omega$ of the lowest supported mode (higher frequency modes are disregarded or rather not considered explicitly). The particles interact with the cavity mode and with each other through the columbic interaction and, for simplicity, are assumed spinless. The corresponding Hamiltonian has the following form:

$$\hat{H} = \frac{1}{2m}[\hat{p} - e\hat{A}(\vec{r})]^2 + \sum_i \frac{1}{2m_i}[\hat{p}_i - Z_i e\hat{A}(\vec{r}_i)]^2 + V_{coul}(\vec{r}, \{\vec{r}_i\}) + \hbar\omega\hat{a}^\dagger\hat{a} \qquad (4)$$

where $\hat{a}^\dagger(\hat{a})$ denote the creation( annihilation) operator of the cavity mode, e, $m$, $\hat{p}$ and $\vec{r}$ are the excess electron's charge, mass, momentum and position while $Z_i e$, $m_i$, $\hat{p}_i$, $\vec{r}_i$ are the charge, mass, momentum and position of the $i$-particle respectively. $V_{coul}(\vec{r}, \{\vec{r}_i\})$ is the electrostatic interaction between all charged particles, $Z_i$ is a charge of $i$-particle and

$$\hat{A}(\vec{r}) = \sqrt{\frac{\hbar}{2\omega\Omega\varepsilon_0}} \vec{\xi}\{\exp(-i\vec{k}\vec{r})\hat{a}^\dagger + \exp(i\vec{k}\vec{r})\hat{a}\} \qquad (5)$$

is the vector potential of the electromagnetic field. Here we have assumed for simplicity that a single cavity mode dominates the investigated process, but a sum over relevant cavity modes could be taken as well. In Eq. (5) $\varepsilon_0$ is the vacuum permittivity while $\vec{\xi}$ and $\vec{k}$ denote the mode polarization and wave vectors. Assuming a transverse field, these vectors are orthogonal to each other, $\vec{\xi}\cdot\vec{k}=0$. Generally, the electrostatic interaction $V_{coul}(\vec{r},\{\vec{r}_i\})$ is affected by the cavity environment[36,67] which can be important for obtaining eigenstates of the molecular Hamiltonian. In the present discussion these states are assumed to be known.

Considerable simplification is achieved when the characteristic size of the molecular system, denoted $d_s$, is assumed to be much smaller than the mode wavelength. Under this assumption, $|kd_s| \ll 1$, we can put $\hat{A}(\vec{r}) \approx \hat{A}(0)$:

$$\hat{A}(0) = \sqrt{\frac{\hbar}{2\omega\Omega\varepsilon_0}} \vec{\xi}\{\hat{a}^\dagger + \hat{a}\} \qquad (6)$$

The molecular system (excluding the excess electron) is characterized by the total dipole moment

$$\vec{\mu}_s = \sum_n Z_n e \vec{r}_n. \qquad (7)$$

Since the system is neutral ($\sum_n Z_n = 0$), $\vec{\mu}_s$ does not depend on the choice of origin of coordinates.

Next, we perform a unitary transformation

$$\hat{U} = \exp\left[-\frac{i}{\hbar}\{\vec{r}e + \vec{\mu}_s\}\hat{A}(0)\right] \qquad (8)$$

in order to eliminate the vector potential from Eq. (4). This leads to (Appendix A)

$$\hat{H}' \equiv \hat{U}\hat{H}\hat{U}^\dagger = \frac{\hat{p}^2}{2m} + \sum_n \frac{1}{2m_n}\hat{p}_n^2 + V_{coul}(\vec{r},\{\vec{r}_i\}) + \hbar\omega\hat{a}^\dagger\hat{a} - (\vec{r}e + \vec{\mu}_s)\cdot\hat{\mathcal{E}} + \{(\vec{r}e + \vec{\mu}_s)\cdot\vec{\xi}\}^2 \frac{1}{2\Omega\varepsilon_0} \qquad (9)$$

where $\hat{\mathcal{E}} = i\vec{\xi}\sqrt{\frac{\hbar\omega}{2\Omega\varepsilon_0}}(\hat{a} - \hat{a}^\dagger)$ is the operator representing the electric field associated with the corresponding cavity mode. The field-matter interaction in the transformed Hamiltonian (9) now has the familiar dipole-field interaction form. In addition, the emerging self-interaction term $(2\Omega\varepsilon_0)^{-1}[\{\vec{r}e+\vec{\mu}_s\}\vec{\xi}]^2$ will be seen to play a crucial role in our model. In what follows we will regroup the first three terms representing the material system as $\hat{H}_M$ - the molecular Hamiltonian, $\hat{H}_B$ - a bath (environment) and $\hat{H}_{MB}$ their mutual interaction, yielding the form

$$\hat{H} = \hat{H}_M + \hat{H}_B + \hat{H}_{MB} + \hbar\omega\hat{a}^\dagger\hat{a} - (e\vec{r}+\vec{\mu}_s)\cdot\vec{\mathcal{E}} + \{(e\vec{r}+\vec{\mu}_s)\cdot\vec{\xi}\}^2 \frac{1}{2\Omega\varepsilon_0} \quad (10)$$

Next, we employ the model given by Eq. (10) to the case of a molecular system with an extra electron put on an unoccupied molecular orbital and the whole material system is placed inside an electromagnetic cavity-resonator. The excess electron is assumed to move between orbitals of the molecular system, taken to be orthogonal. To focus on the standard model used to describe molecular electron transfer we further take $\hat{H}_M + \hat{H}_B + \hat{H}_{MB}$ to represent a 2-state molecular model and a harmonic bath with the standard polaronic interaction:

$$\hat{H}_M + \hat{H}_B + \hat{H}_{MB} \to E_D|D\rangle\langle D| + \{E_A + \sum_j \lambda_j(\hat{b}_j^\dagger + \hat{b}_j)\}|A\rangle\langle A| + H_{DA}|D\rangle\langle A| + H_{AD}|A\rangle\langle D| + \sum_j \hbar\nu_j \hat{b}_j^\dagger \hat{b}_j$$

(11)

Here $|D\rangle$ and $|A\rangle$ stand for states with the excess electron located on donor and the acceptor sites, respectively, $\hat{b}_j^\dagger(\hat{b}_j)$ denotes creation (annihilation) operator of a vibrational mode of frequency $\nu_j$ associated with the bath or an inter/intra molecular vibration and $\lambda_j$ is the vibronic coupling parameter.

Next, consider the last two terms in Eq. (10), the field-molecular system coupling and the self-interaction, that involve the operators $\{e\vec{r}+\vec{\mu}_s\}$ and $\{(e\vec{r}+\vec{\mu}_s)\cdot\vec{\xi}\}^2$. These operators will enter our rate calculations via matrix elements involving the system states $|A\rangle$ and $|D\rangle$ which are assumed to constitute a complete basis for the electronic subspace of this electron transfer problem. We further assume that the neutral molecular system does not have a permanent dipole moment, so

$$\vec{d}_{nl} \equiv \langle n|-\{e\vec{r}+\vec{\mu}_s\}|l\rangle = -\langle n|e\vec{r}|l\rangle \quad (12a)$$

$$\varepsilon_{nl} \equiv \langle n|\{(e\vec{r}+\vec{\mu}_s)\cdot\vec{\xi}\}^2|l\rangle = \langle n|\{e\vec{r}\cdot\vec{\xi}\}^2|l\rangle \quad (12b)$$

where $n$ and $l$ stand for these D and/or A states. We denote the matrix elements on the RHS of Eq. (12a) by

$$\langle D|\vec{r}|D\rangle \equiv -\vec{d}_{DD}/e \tag{13a}$$

$$\langle A|\vec{r}|A\rangle \equiv -\vec{d}_{AA}/e \tag{13b}$$

which are essentially the donor and acceptor positions, $-\vec{d}_{DD}/e = \vec{r}_D$ and $-\vec{d}_{AA}/e = \vec{r}_A$ respectively, and

$$\vec{d}_{DA} = \vec{d}_{AD}^* = \langle D|-e\vec{r}|A\rangle \tag{14}$$

- the transition dipole moment between the donor and acceptor orbitals. Obviously, the magnitude of the latter depends on the overlap between these orbitals and is of the order of the tunneling matrix element that determines the electron transfer rate. In terms of these quantities, the matrix elements of Eq. (12b) take the form

$$\varepsilon_{DD} = \langle D|\{e\vec{r}\cdot\vec{\xi}\}^2|D\rangle = \langle D|\{e\vec{r}\cdot\vec{\xi}\}|D\rangle\langle D|\{e\vec{r}\cdot\vec{\xi}\}|D\rangle + \langle D|\{e\vec{r}\cdot\vec{\xi}\}|A\rangle\langle A|\{e\vec{r}\cdot\vec{\xi}\}|D\rangle$$
$$(\vec{d}_{DD}\cdot\vec{\xi})^2 + |\vec{d}_{DA}\cdot\vec{\xi}|^2 \tag{15a}$$

$$\varepsilon_{AA} = \langle A|\{e\vec{r}\vec{\xi}\}^2|A\rangle = \langle A|\{e\vec{r}\cdot\vec{\xi}\}|A\rangle\langle A|\{e\vec{r}\cdot\vec{\xi}\}|A\rangle + \langle A|\{e\vec{r}\cdot\vec{\xi}\}|D\rangle\langle D|\{e\vec{r}\cdot\vec{\xi}\}|A\rangle$$
$$= (\vec{d}_{AA}\cdot\vec{\xi})^2 + |\vec{d}_{DA}\cdot\vec{\xi}|^2 \tag{15b}$$

$$\varepsilon_{DA} = \langle D|\{e\vec{r}\cdot\vec{\xi}\}^2|A\rangle = \langle D|\{e\vec{r}\cdot\vec{\xi}\}|D\rangle\langle D|\{e\vec{r}\cdot\vec{\xi}\}|A\rangle + \langle D|\{e\vec{r}\cdot\vec{\xi}\}|A\rangle\langle A|\{e\vec{r}\cdot\vec{\xi}\}|A\rangle =$$
$$= (\vec{d}_{DA}\cdot\vec{\xi})\{(\vec{d}_{DD}+\vec{d}_{AA})\cdot\vec{\xi}\} \tag{15d}$$

Using Eq. (12-15) we can rewrite the Hamiltonian (11) in the following form

$$\hat{H} = E_D|D\rangle\langle D| + \left\{E_A + \sum_j \lambda_j\left(\hat{b}_j^\dagger + \hat{b}_j\right)\right\}|A\rangle\langle A| + H_{DA}|D\rangle\langle A| + H_{AD}|A\rangle\langle D| + \sum_j \hbar\nu_j \hat{b}_j^\dagger \hat{b}_j$$
$$+\hbar\omega\hat{a}^\dagger\hat{a} + \hbar\omega g_D(\hat{a}-\hat{a}^\dagger)|D\rangle\langle D| + \hbar\omega g_A(\hat{a}-\hat{a}^\dagger)|A\rangle\langle A| + \hbar\omega|g_D|^2|D\rangle\langle D| + \hbar\omega|g_A|^2|A\rangle\langle A| +$$
$$\hbar\omega(\hat{a}-\hat{a}^\dagger)(t_{DA}|D\rangle\langle A| + t_{AD}|A\rangle\langle D|) + \hbar\omega|t_{DA}|^2 - \hbar\omega(g_D+g_A)(t_{DA}|D\rangle\langle A| + t_{AD}|A\rangle\langle D|)$$
$$\tag{16}$$

where

$$g_D = i\sqrt{\frac{1}{2\hbar\omega\Omega\varepsilon_0}}\vec{d}_{DD}\cdot\vec{\xi} \tag{17a}$$

$$g_A = i\sqrt{\frac{1}{2\hbar\omega\Omega\varepsilon_0}}\vec{d}_{AA}\cdot\vec{\xi} \tag{17b}$$

$$g_{DA} = -g_{AD} = g_D - g_A \tag{17c}$$

$$t_{DA} = -t^*_{AD} = i\sqrt{\frac{1}{2\hbar\omega\Omega\varepsilon_0}}\vec{\xi}\cdot\vec{d}_{DA} \tag{17d}$$

The first four terms in the Hamiltonian (16) correspond to the standard electron transfer process. The other terms represent the effect of coupling to a cavity mode that may potentially become important when the cavity volume $\Omega$ is small. Another useful form of this Hamiltonian can be obtained by making the unitary (polaron-type) transformation (see Appendix B):

$$\hat{U} = \exp\{(g_D^*\hat{a}^\dagger - g_D\hat{a})|D\rangle\langle D| + (g_A^*\hat{a}^\dagger - g_A\hat{a})|A\rangle\langle A|\} \tag{18}$$

which transform the Hamiltonian (16) into the following from (Appendix C):

$$\hat{H}' = \hat{U}\hat{H}\hat{U}^\dagger = E_D|D\rangle\langle D| + \left\{E_A + \sum_j \lambda_j\left(\hat{b}_j^\dagger + \hat{b}_j\right)\right\}|A\rangle\langle A| + \sum_j \hbar\nu_j\hat{b}_j^\dagger\hat{b}_j$$
$$+\hbar\omega\hat{a}^\dagger\hat{a} + \tilde{H}_{DA}|D\rangle\langle A|\exp\{g_{DA}^*\hat{a}^\dagger - g_{DA}\hat{a}\} + \tilde{H}_{AD}|A\rangle\langle D|\exp\{g_{AD}^*\hat{a}^\dagger - g_{AD}\hat{a}\}) + \tag{19}$$
$$+\hbar\omega\{t_{DA}(\hat{a}-\hat{a}^\dagger)|D\rangle\langle A|\exp\{g_{DA}^*\hat{a}^\dagger - g_{DA}\hat{a}\} + t_{AD}\exp\{g_{AD}^*\hat{a}^\dagger - g_{AD}\hat{a}\}|A\rangle\langle D|(\hat{a}-\hat{a}^\dagger)\}$$

where $\tilde{H}_{AD} = H_{AD} + \hbar\omega t_{AD}g_{AD}$.

Several observations on the physical contents of this Hamiltonian can be made at the outset:

(a) As seen explicitly in the form (19), the Hamiltonian depends only on the relative distance between the donor and the acceptor. This is of course an expected result, however note that to obtain it was important to keep the self-interaction (last) term in Eq. (9).

(b) If $t_{AD} = 0$ Eq. (16) and (19) are reduced the standard spin-boson model. Alternatively, by disregarding coupling to the boson $\{\hat{b}_j, \hat{b}_j^\dagger\}$ field by putting $g_{AD} = 0$ we recover the standard Jaynes-Cummings model[68] of coupled two level system and Harmonic mode.

(c) Viewing this Hamiltonian from the perspective of the electron transfer problem, we notice that in addition to the "standard" coupling terms associated with the non-adiabatic coupling $H_{DA}$ we encounter coupling between the donor and acceptor states arising from their coupling to the common cavity modes and characterized by the transition dipole coupling $t_{DA}$, Eq. (17d). In the next Section we discuss the quantitative implications of this interaction.

(d) We expect that this effect of coupling to the cavity mode will depend on the relative characteristic times, $\tau_e$, of the electron transfer and $\tau_c = \omega^{-1}$ of the cavity dynamics. It should be emphasized, however, that the relevant characteristic time for the electron transfer process is not the observed rate of electron transfer (which is usually dominated by the underlying nuclear dynamics) but by the time of actual electronic charge reorganization during a tunneling event, the so called tunneling time[69,70], over which the electronic charge changes from being localized near the donor to being near the acceptor. Depending on the tunneling barrier, this time may be of order 0.1-1 fs[71]. The case where it is of the order of the cavity mode period requires a reconsideration of the molecular electronic structure in the presence of the cavity mode[36]. Simpler general statements can be made in the limits where this time is short or long or comparable to this period, which we consider next.

*Fast electron, slow cavity mode.* Consider first the case $\tau_e \ll \tau_c$, that is, electron tunneling is instantaneous on the timescale of the cavity mode. The rate of the tunneling event depends in this case only on the initial state of the cavity mode (which, just as nuclear states, remains frozen during the tunneling event). In this case the effect of the cavity mode on the electron transfer process will be the same as other slow modes associated with intramolecular or environmental nuclear motions. One caveat in this consideration is that the states of the cavity mode should be calculated in the presence of the molecular system, however, since the molecule is much smaller than the cavity, a mode delocalized in the cavity is only slightly affected by the molecule presence, and we assume that its lowest states are well approximated by those of the free cavity mode.

The calculation of the cavity effect on the transfer rate than becomes analogous to that of the other slow bosons $\{\hat{b}_j, \hat{b}_j^\dagger\}$. To evaluate the transfer rate in this limit we perform the polaron transformation also with respect to the vibrational motions[72]. The Hamiltonian (19) transforms to

$$\hat{H} = E_D|D\rangle\langle D| + \left\{E_A - \sum_j \frac{\lambda_j^2}{\hbar\nu_j}\right\}|A\rangle\langle A| + \sum_j \hbar\nu_j \hat{b}_j^\dagger \hat{b}_j + \hbar\omega \hat{a}^\dagger \hat{a} +$$

$$\tilde{H}_{DA}|D\rangle\langle A|\exp\{g_{DA}^*\hat{a}^\dagger - g_{DA}\hat{a}\}\prod_j \exp\left\{\frac{\lambda_j}{\hbar\nu_j}(\hat{b}_j^\dagger - \hat{b}_j)\right\} +$$

$$\tilde{H}_{AD}|A\rangle\langle D|\exp\{g_{AD}^*\hat{a}^\dagger - g_{AD}\hat{a}\}\prod_j \exp\left\{-\frac{\lambda_j}{\hbar\nu_j}(\hat{b}_j^\dagger - \hat{b}_j)\right\} + \quad (18)$$

$$\hbar\omega\{t_{DA}(\hat{a} - \hat{a}^\dagger)|D\rangle\langle A|\exp\{g_{DA}^*\hat{a}^\dagger - g_{DA}\hat{a}\}\prod_j \exp\left\{\frac{\lambda_j}{\hbar\nu_j}(\hat{b}_j^\dagger - \hat{b}_j)\right\} +$$

$$t_{AD}\exp\{g_{AD}^*\hat{a}^\dagger - g_{AD}\hat{a}\}|A\rangle\langle D|(\hat{a} - \hat{a}^\dagger)\prod_j \exp\left\{-\frac{\lambda_j}{\hbar\nu_j}(\hat{b}_j^\dagger - \hat{b}_j)\right\}$$

Note that only the vibrational reorganization energy $E_R \equiv \sum_j \frac{\lambda_j^2}{\hbar\nu_j}$ contributes to the renormalization of the donor-acceptor energy gap, while the cavity mode does not contribute. This is because we kept the self-interaction dipole term in the Hamiltonian (9). However, the presence of the shift operator $\exp\{g_{AD}^*\hat{a}^\dagger - g_{AD}\hat{a}\}\prod_j \exp\left\{\lambda_j\left(\hat{b}_j^\dagger + \hat{b}_j\right)\right\}$ implies that matrix elements between donor and acceptor states will be dressed by matrix (Franck-Condon-type) elements of the shift operators between the associated initial and final states of the phonon and photon states. Denoting by $\{n\}(=n_1,...,n_j,...)$ and $\{n'\}$ the initial and final states of the vibrational environment and by and by $m$ and $m'$ initial and final states of the cavity mode, a typical coupling matrix element is

$$V_{DA}^{\{n\}\{n'\}mm'} = \langle D\,m\,\{n\}|\hat{H}|A\,m'\,\{n\}'\rangle = F_{\{n\}\{n'\}}G_{DA}^{mm'} \quad (21)$$

where

$$F_{\{n\}\{n'\}} = \prod_j F_{n_j,n_j'}\left(\frac{\lambda_j}{\hbar\nu_j}\right) \quad (22a)$$

$$G_{DA}^{mm'} = \{H_{DA} - \hbar\omega t_{DA}g_{DA}\}F_{mm'}(g_{AD}) + \hbar\omega t_{DA}\{F_{m+1,m'}(g_{AD}) - F_{m-1,m'}(g_{AD})\} \quad (22b)$$

and where $F_{kl}(x)$ are matrix elements of shift operators (whose squares are Franck – Condon factors):

$$F_{kl}(x) = \langle k | \hat{F}(x) | l \rangle = \sqrt{\frac{k!}{l!}} (x)^{l-k} \exp(-\frac{1}{2}|x|^2) L_k^{l-k}(|x|^2), \qquad (22c)$$

with $L_k^{l-k}$ denoting Laguerre polynomials.

Given the coupling (22) the electron transfer rate can be calculated in the standard way from the Golden rule (see, e.g., chapter 16 of Ref. 72)). The effect of coupling to the cavity mode can be seen explicitly by considering the Marcus classical limit for the vibrational contribution. For the rate of transition from an initial molecule-cavity state $|Dm\rangle$ to a final state $|Am'\rangle$ this yields

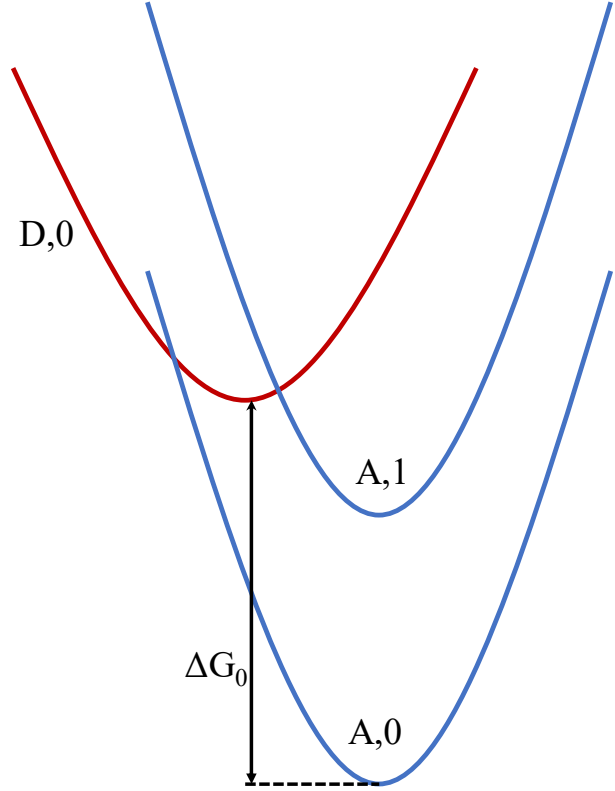

Figure 1. A scheme of the cavity-modified ET process. In the initial state the electron is on the donor and the cavity field is in its ground state. In the final states the electron is on the acceptor and the cavity mode may be in any state consistent with energy conservation.

$$\kappa_{Dm \to Am'} = \sqrt{\frac{\pi}{\hbar^2 k_B T E_R}} \left| G_{DA}^{mm'} \right|^2 \exp\left\{ -\frac{(E_D - E_A - E_R - \hbar\omega\Delta m)^2}{4 k_B T E_R} \right\} \qquad (23)$$

while the total rate of electron transfer starting from state $|Dm\rangle$ is

$$\kappa_{D \to A} = \sum_m P_m \sum_{m'} \kappa_{Dm \to Am'}. \qquad (24)$$

Here $\Delta m = m' - m$ and $P_m$ is the thermal (Bose – Einstein with possible corrections for a lossy cavity) population of the cavity mode. If the cavity-mode is initially in the ground state, $m = 0$, it is obvious from Fig. 1 that the rate is dominated by the $m=0 \rightarrow m'=0$ transition in the normal Marcus regime, however transitions to higher $m'$ states may become important when the standard process outside the cavity is in the inverted regime. Possible implications are discussed in the next Section.

*Slow electron, fast cavity mode.* Next consider the opposite case $\tau_e \gg \tau_c$, namely the tunneling event is much slower than the characteristic cavity dynamics. In this case we can use the Born-Oppenheimer approximation with the molecular electronic and (obviously) nuclear degrees of freedom assumed slow compared with the cavity dynamics. Accordingly, we look for eigenfunctions and eigenvalues of the Hamiltonian (9) of the form

$$\psi = \psi_{el,m}(\vec{r},\vec{Q})\left|m;\vec{r},\vec{Q}\right\rangle \qquad (24)$$

where $\left|m;\vec{r},\vec{Q}\right\rangle$ are eigenstates (characterized by the quantum number $m$) of the cavity Hamiltonian, obtained from the Hamiltonian (9) by excluding the kinetic energy operators of the molecular motions,

$$\hat{H}_c = \hbar\omega\hat{a}^\dagger\hat{a} - ie\sqrt{\frac{\hbar\omega}{2\Omega\varepsilon_0}}\vec{r}\cdot\vec{\xi}(\hat{a}-\hat{a}^\dagger) + (e\vec{r}\cdot\vec{\xi})^2\frac{1}{2\Omega\varepsilon_0} + V(\vec{r};\vec{Q}) \qquad (25)$$

This Hamiltonian describes the cavity for a given instantaneous molecular vibronic configuration – the excess electron position $\vec{r}$ and the nuclear configuration $\vec{Q}$. It may be rewritten in the form

$$\hat{H}_c = V(\vec{r};\vec{Q}) + \hbar\omega(\hat{a}^\dagger + gx)(\hat{a} + g^*x) \qquad (26)$$

where $g = ie/\sqrt{2\varepsilon_0\Omega\hbar\omega}$ and $x = \vec{r}\cdot\vec{\xi}$. The eigenfunctions and eigenvalues of this Hamiltonian are

$$\left|m;\vec{r},\vec{Q}\right\rangle = \left|n;x\right\rangle = \exp\{x(g^*\hat{a}^\dagger - g\hat{a})\}\left|m\right\rangle \qquad (27)$$

which are linearly shifted eigenstates $\left|m\right\rangle$ of the free cavity Hamiltonian $\hbar\omega\hat{a}^\dagger\hat{a}$. The corresponding eigenvalues are

$$E_m(\vec{r};\vec{Q}) = m\hbar\omega + V(\vec{r};\vec{Q}) \qquad (28)$$

The shift in the mode wavefunction depends on the excess electron position, however in contrast to standard applications of the Born-Oppenheimer approximation, the eigenvalues depend on the state of the slow subsystem (here the vibronic configuration) only via a purely additive term. These eigenvalues constitute the adiabatic potential energy surfaces for the slow molecular (electronic and nuclear) dynamics, namely the Schrödinger equation for the wavefunction $\psi_{el,m}(\vec{r},\vec{Q})$ of the (slow) molecular system is

$$\left(\frac{\hat{p}^2}{2m} + E_m(\vec{r};\vec{Q})\right)\psi_{el,n}(\vec{r};\vec{Q}) = E_{tot}\psi_{el,m}(\vec{r};\vec{Q}) \tag{29}$$

where $E_{tot}$ is the energy of the total molecule-field system. However, since the field term $m\hbar\omega$ does not depend on the molecular configuration, it follows that the molecular wavefunction does not depend, in this limit, on the state of the cavity mode, and satisfies the free molecule Schrödinger equation

$$\left(\frac{\hat{p}^2}{2m} + V(\vec{r};\vec{Q})\right)\psi_{el}(\vec{r};\vec{Q}) = E_{el}\psi_{el}(\vec{r};\vec{Q}) \tag{30}$$

so that $E_{tot} = E_{el} + m\hbar\omega$. Thus, in this limit of fast cavity dynamics and slow electron, the electron dynamics, in particular the electron transfer process, is not affected by the cavity. The mode instantaneously adjusts to the electron motion and electron transfer is unperturbed by the field. In this case the non-adiabatic couplings between dressed states can be obtained from Eq. (22b) by putting $F_{mm'}(x) = \delta_{mm'}$:

$$V_{DA}^{\{n\}\{n'\}mm'} = F_{\{n\}\{n'\}}H_{DA}\delta_{mm'} + \hbar\omega t_{DA}F_{\{n\}\{n'\}}\left\{\delta_{m+1,m'} - \delta_{m-1,m'}\right\} \tag{31}$$

The last term in (31) corresponds to the standard dipole coupling between the dressed states. The expression for the rate in this limit is similar to Eq. (23) where the corresponding coupling elements $G_{DA}^{mm'}$ is now given by $G_{DA}^{mm'} = H_{DA}\delta_{mm'} + \hbar\omega t_{DA}\left\{\delta_{m+1,m'} - \delta_{m-1,m'}\right\}$. The cavity environment can affect electron transfer only when it is accompanied by a change in the photon population. If the cavity-mode is initially unpopulated this again can happen in the inverted Marcus regime. We return to this issue in Section 4.

The two limits considered above can also be easily understood by analyzing the time dependent dynamics in the Heisenberg picture (see Appendix D).

*Cooperative effects.* The strong coupling of many molecular systems to an optical mode delocalized within the cavity is known to induce cooperative molecular response[44,53,55,73]. Manifestation of such effects for the vibronic dynamics in many-molecule systems will be studied separately, and here we only note its simplest possible realization. We start by rewriting the Hamiltonian (19) for the case of an ensemble of non-interacting donor-acceptor pairs:

$$\hat{H} = \sum_{k=1}^{N} E_D^k |D^k\rangle\langle D^k| + \left\{ E_A^k + \sum_j \lambda_j^k \left[ \hat{b}_{j,k}^\dagger + \hat{b}_{j,k} \right] \right\} |A^k\rangle\langle A^k| + \sum_j \hbar \nu_{j,k} \hat{b}_{j,k}^\dagger \hat{b}_{j,k}$$
$$+ \tilde{H}_{DA,k} |D^k\rangle\langle A^k| \exp\{g_{DA,k}^* \hat{a}^\dagger - g_{DA,k} \hat{a}\} + \tilde{H}_{AD,k} |A^k\rangle\langle D^k| \exp\{g_{AD,k}^* \hat{a}^\dagger - g_{AD,k} \hat{a}\} +$$
$$+ \hbar \omega \hat{a}^\dagger \hat{a} + \tag{32}$$
$$\hbar \omega \{ t_{DA,k} (\hat{a} - \hat{a}^\dagger) |D^k\rangle\langle A^k| \exp\{g_{DA,k}^* \hat{a}^\dagger - g_{DA,k} \hat{a}\}$$
$$+ t_{AD,k} \exp\{g_{AD,k}^* \hat{a}^\dagger - g_{AD,k} \hat{a}\} |A^k\rangle\langle D^k| (\hat{a} - \hat{a}^\dagger) \}$$

Here, parameters $g_{DA,k}$ and $t_{DA,k}$ are defined as in Eqs (13)-(17) for the different molecules $k$. Assuming that the pairs are identical, the spatial size of the system is much smaller than the mode wavelength and the pairs are aligned in the same direction with respect to the polarization of the cavity field, we have $d_{XY,k} = d_{XY}$, $g_{XY,k} = g_{XY}$ and $t_{XY,k} = t_{XY}$ for all $X, Y = A, B$. In this case, the Hamiltonian (34) represents a generalized version of the Dicke (or Tavis-Cummings) model[74] where, instead of spontaneous emission $|e0\rangle \to |g1\rangle$, a cavity photon(s) can be produced through the electron transfer: $|Dm\rangle \to |Am'\rangle$ with $m' = m+1$. Limiting the present discussion to the case $m=0$ and $m'=1$ we can use the result of the Dicke model:

$$\kappa_{D0 \to A1, N(n)} = \frac{(N/2+n)(N/2-n+1)}{N} \kappa_{D \to A1} \tag{33}$$

where $\kappa_{D \to A, N(n)}$ is the transfer rate for the process $|D0\rangle \to |A1\rangle$ in a system that has started with $N_D = N$ molecules in the donor state ($N_A = N - N_D = 0$), namely $n \equiv (N_D - N_A)/2 = N/2$ and has reached a state with a smaller $n$. Initially when $n = N/2$ we have $\kappa_{D0 \to A1, N_D = N} = \kappa_{D0 \to A1,1}$, i.e. the single molecule rate. When $m=0$ (the half of all electrons have been transferred) the cavity induced rate becomes

$$\kappa_{D0 \to A1, N_D = N/2} = \frac{1}{4}(N+2)\kappa_{D0 \to A1,1} \tag{34}$$

that is, enhances by a factor of $(N+2)/4$. These results, which follows those of the Dicke model[55,74,75], should be regarded as tentative: First, in the present case, unlike in the Dicke model, the process $D0 \to A0$ can occur simultaneously with $D0 \to A1$ so that the value of $n$ does not reflect the number of photons that have been emitted. Second, the nuclear motion which is the driving force of the non-adiabatic ET should also manifest itself as a collective mode which is (perhaps) achievable at low temperatures. Finally, the lifetime of the cavity mode (i.e. the lifetime of mixed polaritonic states) should be much larger than the inverse ET rate.

## 3. Results and Discussion

In this section we examine the implications of the results obtained above, focusing on the two extreme limits were the electron tunneling time is fast or slow compare with the cavity characteristic time (inverse mode frequency). To determine the parameters to be used in our estimates we first note that for the present problem (at least without considering collective effects) the quality factor of the cavity is of secondary importance, so we may consider plasmonic cavities where the field can be strongly focused. When dominated by a single plasmon/cavity mode of frequency $\omega$, the local field operator has the form

$$\hat{\mathcal{E}} = iM\left(\hat{a}_\omega - \hat{a}_\omega^\dagger\right) \tag{35}$$

where $\hat{a}_\omega$ and $\hat{a}_\omega^\dagger$ are the creation and annihilation operators for this mode. The parameter $M$ depends on cavity properties and is sometimes expressed in terms of an effective cavity volume $\Omega$ using the expression for a vacuum space between two plane reflectors, $M = \sqrt{\hbar\omega/2\varepsilon_0\Omega}$. M can be calculated for simple structures or estimated from observed single molecule exciton-plasmon Rabi splitting $\Lambda$. Below we show results for two parameter sets associated with different cavities. First, recent reports[76–80] indicate that a single molecule splitting of order $\Lambda \sim 100$meV can be achieved in a plasmonic cavity with $\omega \sim 2$eV and a molecular transition dipole of order $\mu \sim$ 4D, implying a local field of order $M = \Lambda/2\mu \approx 0.57 \times 10^9 \frac{V}{m}$, which corresponds to an effective cavity volume $\sim 40 nm^3$. We refer the system characterized by these parameters as Cavity A. Second, for the cavity that forms between a doped-InSb sphere of radius 10nm separated by 1nm from a planar surface of the same substance, which supports an infrared plasmon resonance at $\omega = 0.2 eV$, we have found[81] $M \approx 0.1 \times 10^9 \frac{V}{m}$, corresponding to an effective cavity volume $\sim 160 nm^3$. This structure is referred to below as cavity B.

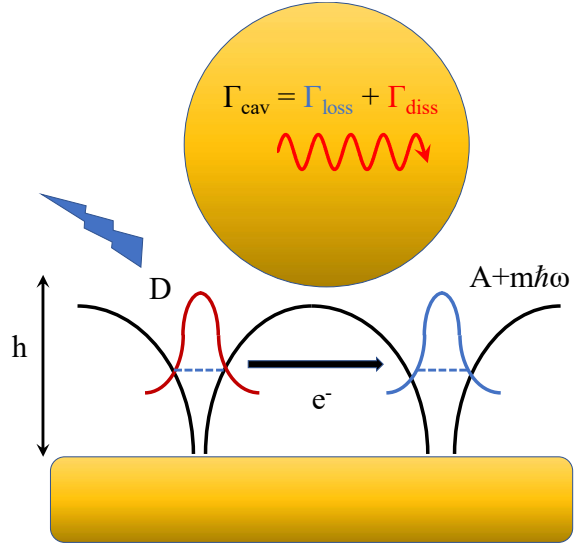

Figure 2. A schematic representation of the model system: a donor-acceptor pair is put into a plasmonic nanocavity. The reaction coordinate is aligned along the polarizability vector.

Next, consider the parameters that characterize the electron transfer process. The matrix elements $H_{DA}$ and $t_{DA}$ needed to calculate G (Eq. (22b)) and its limiting forms can be roughly estimated in terms of the barrier $\Delta E$ to electron tunneling, the donor-acceptor distance $r_{DA}$ and the electron mass $m_e$ following the arguments advanced by Mulliken (see Eqs. 4, 5 and 16 in Ref 83). We first consider the overlap between donor-localized and acceptor-localized orbitals calculated from their exponentially decaying tails in the barrier region

$$S \sim \frac{\sqrt{2m_e \Delta E} r_{DA}}{\hbar} \exp\left(-\frac{\sqrt{2m_e \Delta E} r_{DA}}{\hbar}\right) \qquad (36)$$

Using S, $H_{DA}$ and $t_{DA}$ can be estimated given the value of M and using[82,83] $H_{DA} \sim S\Delta E$, $d_{DA} \sim S r_{DA} e / 2$ and $t_{DA} = M d_{DA}$.

Consider first the case of slow electron and fast cavity mode, using the parameters of cavity A and taking the tunneling barrier, donor acceptor distance and reorganization energy to be $\Delta E \sim 1 eV$, $r_{DA} \sim 1 nm$ and $E_R = 1$ eV, respectively. An estimate (based on a square barrier model[84,85]) for the tunneling time is $\tau = \frac{r_{DA}}{\sqrt{2\Delta E m_e^{-1}}} \approx 1.7 fs$, so this limit is indeed approached for $\hbar\omega \geq 2 eV$. As discussed above, in this limit the electronic energy surfaces do not change, except for a possible vertical shift associated with dressing by cavity photons. However, the coupling element $t_{DA}$, Eq. (17d), can induce inelastic tunneling, where electron transfer is accompanied by photon generation or annihilation. In particular, if the cavity mode is initially unpopulated, two terms in Eq.(23) will contribute to the total rate:

$$\kappa_{D \to A} = \sqrt{\frac{\pi}{\hbar^2 k_B T E_R}} \left( |H_{DA}|^2 \exp\left\{-\frac{(-\Delta G_0 - E_R)^2}{4 k_B T E_R}\right\} + |\hbar \omega t_{DA}|^2 \exp\left\{-\frac{(-\Delta G_0 - E_R - \hbar\omega)^2}{4 k_B T E_R}\right\} \right) \quad (37)$$

The first term in Eq. (37) is the standard Marcus rate where $-\Delta G_0 = E_D - E_A$ stands for the donor-acceptor energy gap. The other represents an inelastic process where electron transfer is accompanied by photon generation and may be significant if $t_{DA}$ is large enough. For a numerical estimate we use the parameter of cavity A together with $r_{DA} = 1nm$ and $\Delta E = 1eV$, which yield $H_{DA} = 245$ cm$^{-1}$ and $t_{DA} = 69$ cm$^{-1}$. The resulting dependence of the electron transfer rate on the donor-acceptor energy gap is shown in Fig. 3. Two peaks are observed, corresponding to the vanishing of the activation energy of the terms on Eq. (37), one at $-\Delta G_0 = E_R$ and the other at $-\Delta G_0 = E_R + \hbar\omega$, where a large $\Delta G_0$ may lead to the generation of a cavity photon. The opposite limit of fast electron tunneling and slow cavity mode may be realized when the tunneling barrier is large, the tunneling distance is small (both implying short tunneling time), and the cavity is constructed to support low frequency plasmons with sufficient local field enhancement. Using the parameters of cavity B, and taking for the tunneling parameters a barrier height of $\Delta E = 3eV$ and donor-acceptor distance $r_{DA} = 1nm$ (corresponding to a square barrier tunneling time ~ $0.98 fs$) puts us in this limit. For these parameters we now get $H_{DA} \approx 30 cm^{-1}$, $|\hbar\omega t_{AD}| \approx 0.5 cm^{-1}$ and, from Eq. (17c), $|g_{AD}| \approx 0.5$ for the optical FC shift parameter. Using also $E_R = 0.2 eV$ for the vibrational reorganization energy and temperature T=300K, the electron transfer rate can be calculated from Eqs. (23) and (24). Figure 4 shows the electron transfer rate from an initial state with an unoccupied cavity mode, using for the sum in Eq. (24) six final states (up to five) photons created during the electron-transfer event):

$$\kappa_{D \to A} = \sqrt{\frac{\pi}{\hbar^2 k_B T E_R}} \times$$
$$\left\{ |H_{DA}|^2 \sum_{m=0}^{5} F_{0m}^2(g_{DA}) \exp\left[-\frac{(-\Delta G_0 - E_R - m\hbar\omega)^2}{4 k_B T E_R}\right] + |\hbar\omega t_{AD}|^2 \sum_{m=1}^{5} F_{1m}^2(g_{DA}) \exp\left[-\frac{(-\Delta G_0 - E_R - m\hbar\omega)^2}{4 k_B T E_R}\right] \right\}$$

(38)

(With the above choice of parameters, $F_{00}^2 = 0.78$ and $F_{05}^2 = 6.4 \times 10^{-6}$, indicating the convergence of this procedure). The red and black curves in Fig. 4 correspond to the standard and the cavity-modified rates, respectively. The main panel shows significant rate enhancements in the inverted region, while the insert shows rate suppression by ~20% in the normal region. The observed behavior reflects the simultaneous occurrence of two effects: First, as before, the observed rate combines optically elastic and inelastic transitions into different final photon states, reflecting the same effect seen in Fig. 3 except that the peaks corresponding to different final photon states now nearly overlap. Additionally, the electron transfer rate is now modulated by an electrodynamic Franck-Condon-type factor associated with the slow cavity mode. The later effect is similar in nature to the effects of inter and intramolecular nuclear motions on electronic transitions - dressing electronic matrix elements with nuclear Franck-Condon factors, and is expected to modify the electron-transfer rates inside infrared cavities.

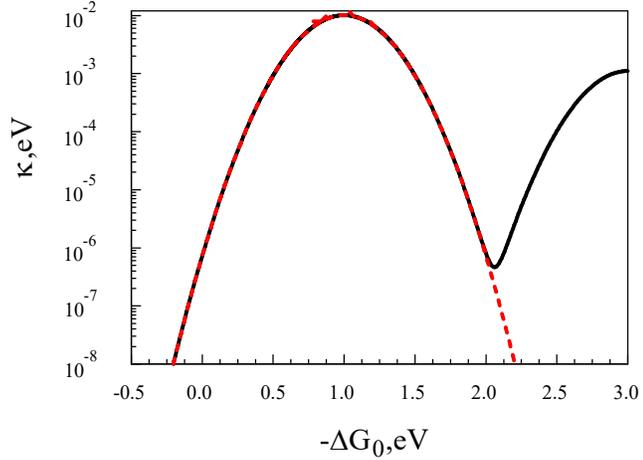

Figure 3. The electron transfer rate obtained from Eq. (37) (fast mode, slow electron limit) plotted against of the donor-acceptor energy gap $\Delta G_0$. The red dashed curve corresponds to the standard (no cavity) rate, first term in (37). The black curve corresponds to the cavity-modified rate. The following parameters are used: $H_{DA}$ = 245 cm$^{-1}$, $\hbar\omega$ =2eV, $\hbar\omega t_{DA}$=69 cm$^{-1}$, $E_R$ = 1.0eV, $T$= 300K. The cavity-modified ET rate (second term in (37) exhibits a second maximum near -$\Delta G_0$= $E_R$+$\hbar\omega$, associated with the crossing of the donor potential surface (D0) and the photon-dressed acceptor surface (A1, see Fig. 1), where the effective coupling is given by $t_{DA}$.

To end this section we note that intermediate situations, where tunneling and cavity dynamics occur on similar timescales, are also expected to show an effect of the cavity environment. Prominent examples of such effects were discussed in Refs. 40-45, where dressing by a cavity mode that bridges the gap between electronic states causes local modifications of potential energy surfaces at the (avoided) intersection of the dressed surfaces.

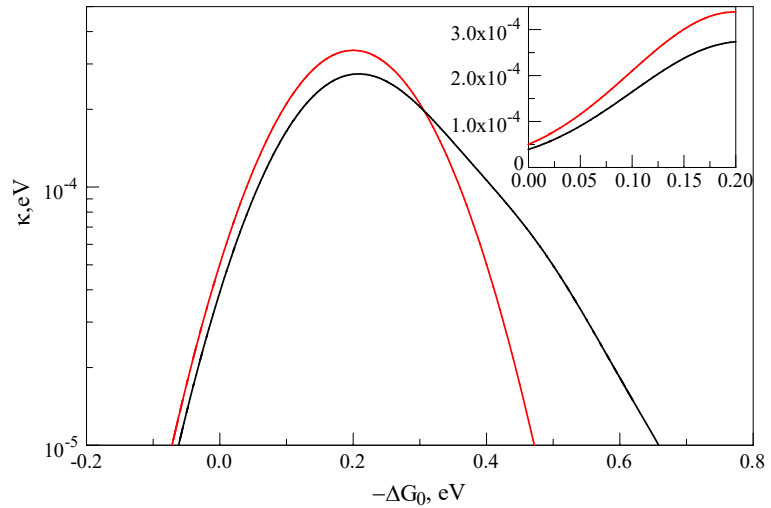

Figure 4. Electron transfer rate as a function of energy gap. The red line presents the cavity-free rate, the black one present to the cavity-modified rate. A log scale is used for the rate axis of the main panel, while the insert is linear in the both axes. The calculations are done in the slow mode limit. Parameters are T=300K, $\hbar\omega$ = 0.2 eV, $H_{DA}$ =30 cm$^{-1}$, $g_{DA}$=0.5, $E_R$ = 0.2eV, the number of FC terms retained in Eq.(38) is 6.

**Conclusions**

We have presented a general framework for describing the interaction between a charged molecular system and a quantized electromagnetic cavity field and applied it to the problem of electron transfer in the presence of such confined field, showing that the transfer rate may be affected by the cavity environment even in the vacuum state of the latter. The effective system Hamiltonian (16) corresponds to a unified spin-boson/ Rabi model where the electromagnetic field induces additional couplings between the donor and the acceptor states that are non-diagonal in the field quantum number and creates additional avoided crossing regions between the dressed electronic states. Two limiting cases, where the cavity mode is slow or fast relative to the characteristic tunneling time, were considered. In both limits the electron transfer rate can be dramatically enhanced in the inverted Marcus regime where electron transfer may be accompanied by the creation of cavity photons. In the limit of slow cavity mode that may be realized in infrared cavities, the rate is renormalized by a Franck-Condon - like factor associated with the overlap between shifted photon energy surfaces. The latter effect translates into a moderate rate reduction in the normal Marcus regime. Our results indicate the potential for controlling electron transfer processes using tunable optical nanocavities.

We have only briefly touched the subject of cooperative behavior. Realization of such behavior in cavity-enhanced molecular electron transfer may require collective response of nuclear motions. Such collective response has been recently demonstrated, even at room temperatures, in infrared cavities[86] and possible implications to electron transfer will be considered in future work.

**ACKNOWLEDGMENTS**

The research of AN is supported by the U.S. National Science Foundation (Grant No. CHE1665291, the Israel-U.S. Binational Science Foundation, the German Research Foundation (DFG TH 820/11-1), and the University of Pennsylvania.

**Appendix A. Dipole gauge transformation**

The Hamiltonian in the dipole gauge is obtained from the form (4) as follows:

$$\hat{H}' = \hat{U}\hat{H}\hat{U}^\dagger = \hat{U}\frac{1}{2m}(\hat{p}-e\hat{A}(0))^2\hat{U}^\dagger + \hat{U}\sum_n \frac{1}{2m_n}(\hat{p}_n - Z_n e\hat{A}(0))^2\hat{U}^\dagger + \hat{U}V_{coul}(\vec{r},\{\vec{r}_i\})\hat{U}^\dagger \quad (A1)$$
$$+\hat{U}\hbar\omega\hat{a}^\dagger\hat{a}\hat{U}^\dagger$$

where $\hat{U}$ is given by Eq.(8). Consider the first term in Eq. (A1):

$$\hat{U}\frac{1}{2m}[\hat{p}-e\hat{A}(0)]^2\hat{U}^\dagger = \hat{U}\frac{1}{2m}(\hat{p}-e\hat{A}(0))\hat{U}^\dagger\hat{U}\frac{1}{2m}(\hat{p}-e\hat{A}(0))\hat{U}^\dagger = \{\hat{U}\frac{1}{2m}(\hat{p}-e\hat{A}(0))\hat{U}^\dagger\}^2 \quad (A2)$$

The term $\hat{U}(\hat{p}-e\hat{A}(0))\hat{U}^\dagger$ needs to be determined:

$$\hat{U}\{\hat{p}-e\hat{A}(0)\}\hat{U}^\dagger = \exp(-\frac{i}{\hbar}\{\vec{r}e + \vec{\mu}_s\}\cdot\hat{A}(0))\{\hat{p}-e\hat{A}(0)\}\exp[\frac{i}{\hbar}\{\vec{r}e + \vec{\mu}_s\}\cdot\hat{A}(0)]$$
$$= \exp(-\frac{i}{\hbar}\{\vec{r}e + \vec{\mu}_s\}\cdot\hat{A}(0))\hat{p}\exp(\frac{i}{\hbar}\{\vec{r}e + \vec{\mu}_s\}\cdot\hat{A}(0)) - e\hat{A}(0)$$
$$= \hat{p} + \exp(-\frac{i}{\hbar}\{\vec{r}e + \vec{\mu}_s\}\cdot\hat{A}(0))\times(-i\hbar\vec{\nabla}\exp\{\frac{i}{\hbar}\{\vec{r}e + \vec{\mu}_s\}\hat{A}(0)\}) - e\hat{A}(0) \quad (A3)$$
$$= \hat{p} + \exp(-\frac{i}{\hbar}\{\vec{r}e + \vec{\mu}_s\}\hat{A}(0))\exp(\frac{i}{\hbar}\{\vec{r}e + \vec{\mu}_s\}\hat{A}(0))e\hat{A}(0) - e\hat{A}(0) = \hat{p}$$

or $\hat{U}\frac{1}{2m}(\hat{p}-e\hat{A}(0))^2\hat{U}^\dagger = \frac{\hat{p}^2}{2m}$ \quad (A4)

And by analogy the second term in Eq. (A1)

$$\hat{U}\sum_n \frac{1}{2m_n}(\hat{p}_n - Z_n e\hat{A}(0))^2 \hat{U}^\dagger = \sum_n \frac{1}{2m_n}\hat{p}_n^2 \tag{A5}$$

The third term in Eq. (A1):

$$\hat{U}V_{coul}(\vec{r},\{\vec{r}_i\})\hat{U}^\dagger = V_{coul}(\vec{r},\{\vec{r}_i\}) \tag{A6}$$

since $V_{coul}(\vec{r},\{\vec{r}_i\})$ and $\hat{U}$ commute.

To evaluate the last term in (A1) we can re-express (10) as follows:

$$\hat{U} = \exp(-\frac{i}{\hbar}\{\vec{r}e + \vec{\mu}_s\}\cdot\hat{\vec{A}}(0)) = \exp(-\frac{i}{\hbar}\sqrt{\frac{\hbar}{2\omega\Omega\varepsilon_0}}\{\vec{r}e + \vec{\mu}_s\}\cdot\vec{\xi}\{\hat{a}^\dagger + \hat{a}\}) \tag{A7}$$

$$= \exp(\lambda^*\hat{a} - \lambda\hat{a}^\dagger)$$

where $\lambda = i\sqrt{\frac{1}{2\hbar\omega\Omega\varepsilon_0}}\{\vec{r}e + \vec{\mu}_s\}\cdot\vec{\xi} \tag{A8}$

The expression (A7) is a shift operator for harmonic oscillator. This means that

$$\hat{U}\hat{a}^\dagger\hat{U}^\dagger = \hat{a}^\dagger + \lambda^* \tag{A9}$$

$$\hat{U}\hat{a}\hat{U}^\dagger = \hat{a} + \lambda \tag{A10}$$

and

$$\hat{U}\hbar\omega\hat{a}^\dagger\hat{a}\hat{U}^\dagger = \hbar\omega\hat{U}\hat{a}^\dagger\hat{U}^\dagger\hat{U}\hat{a}\hat{U}^\dagger = \hbar\omega(\hat{a}^\dagger + \lambda^*)(\hat{a} + \lambda) \tag{A11}$$

Substituting (A4), (A5), (A6) and (A11) in (A1) we get a new expression for the system Hamiltonian:

$$\hat{H}' = \frac{\hat{p}^2}{2m} + \sum_n \frac{1}{2m_n}\hat{p}_n^2 + V_{coul}(\vec{r},\{\vec{r}_i\}) + \hbar\omega\hat{a}^\dagger\hat{a} + \hbar\omega(\lambda^*\hat{a} + \lambda\hat{a}^\dagger) + |\lambda|^2$$

$$= \frac{\hat{p}^2}{2m} + \sum_n \frac{1}{2m_n}\hat{p}_n^2 + V_{coul}(\vec{r},\{\vec{r}_i\}) + \hbar\omega\hat{a}^\dagger\hat{a} - i\{\vec{r}e + \vec{\mu}_s\}\cdot\vec{\xi}\sqrt{\frac{\hbar\omega}{2\Omega\varepsilon_0}}(\hat{a} - \hat{a}^\dagger) + (\{\vec{r}e + \vec{\mu}_s\}\cdot\vec{\xi})^2\frac{1}{2\Omega\varepsilon_0}$$

$$= \frac{\hat{p}^2}{2m} + \sum_n \frac{1}{2m_n}\hat{p}_n^2 + V_{coul}(\vec{r},\{\vec{r}_i\}) + \hbar\omega\hat{a}^\dagger\hat{a} - \{\vec{r}e + \vec{\mu}_s\}\cdot\hat{\vec{\mathcal{E}}} + (\{\vec{r}e + \vec{\mu}_s\}\cdot\vec{\xi})^2\frac{1}{2\Omega\varepsilon_0}$$

$$\tag{A12}$$

**Appendix B. Evaluation of matrix elements of the polaron operator**

We will use the following relations[72]:

$$\hat{A}F(\hat{B}) = F(\hat{B})\hat{A} + [\hat{A},\hat{B}]F'(\hat{B}) \tag{B1}$$

$$\exp(\hat{A}+\hat{B})\exp(1/2[\hat{A},\hat{B}]) = \exp(\hat{A})\exp(\hat{B}) \tag{B2}$$

which are valid if $[[\hat{A},\hat{B}],\hat{A}]=0$ and $[[\hat{A},\hat{B}],\hat{B}]=0$

They give us:

$$\begin{aligned}&\exp\{(g_D^*\hat{a}^\dagger - g_D\hat{a})|D\rangle\langle D|\}\hat{a}|D\rangle\langle D|\exp\{-(g_D^*\hat{a}^\dagger - g_D\hat{a})|D\rangle\langle D|\}\\ &=\exp\{(g_D^*\hat{a}^\dagger - g_D\hat{a})|D\rangle\langle D|\}\exp\{-(g_D^*\hat{a}^\dagger - g_D\hat{a})|D\rangle\langle D|\}(\hat{a}-[\hat{a},g_D^*\hat{a}^\dagger])|D\rangle\langle D|) =\\ &(\hat{a}-g_D^*)|D\rangle\langle D|)\end{aligned} \tag{B3}$$

and

$$\begin{aligned}&\exp\{(g_D^*\hat{a}^\dagger - g_D\hat{a})|D\rangle\langle D|\}|D\rangle\langle A|\exp\{-(g_A^*\hat{a}^\dagger - g_A\hat{a})|A\rangle\langle A|\}\\ &|D\rangle\langle A|(\exp\{(g_D^*-g_A^*)\hat{a}^\dagger - (g_D-g_A)\hat{a})\})\times \exp\{1/2[g_D^*\hat{a}^\dagger - g_D\hat{a}, g_A^*\hat{a}^\dagger - g_A\hat{a}]\}\\ &=|D\rangle\langle A|(\exp\{(g_D^*-g_A^*)\hat{a}^\dagger - (g_D-g_A)\hat{a})\})\times \exp\{1/2(g_D^*g_A - g_D g_A^*)\}\\ &=|D\rangle\langle A|\exp\{g_{DA}^*\hat{a}^\dagger - g_{DA}\hat{a}\}\end{aligned} \tag{B4}$$

From (B3) and (B4) we have

$$\begin{aligned}&\exp\{(g_D^*\hat{a}^\dagger - g_D\hat{a})|D\rangle\langle D|\}\hat{a}|D\rangle\langle A|\exp\{-(g_A^*\hat{a}^\dagger - g_A\hat{a})|A\rangle\langle A|\} =\\ &\exp\{(g_D^*\hat{a}^\dagger - g_D\hat{a})|D\rangle\langle D|\}\hat{a}\exp\{-(g_D^*\hat{a}^\dagger - g_D\hat{a})|D\rangle\langle D|\}\times\\ &\exp\{(g_D^*\hat{a}^\dagger - g_D\hat{a})|D\rangle\langle D|\}|D\rangle\langle A|\exp\{-(g_A^*\hat{a}^\dagger - g_A\hat{a})|A\rangle\langle A|\}\\ &=(\hat{a}-g_D^*|D\rangle\langle D|)|D\rangle\langle A|\exp\{g_{DA}^*\hat{a}^\dagger - g_{DA}\hat{a}\} =\\ &(\hat{a}-g_D^*)|D\rangle\langle A|\exp\{g_{DA}^*\hat{a}^\dagger - g_{DA}\hat{a}\}\end{aligned} \tag{B5}$$

and

$$\begin{aligned}&\exp\{(g_A^*\hat{a}^\dagger - g_A\hat{a})|A\rangle\langle A|\}\hat{a}|A\rangle\langle D|\exp\{-(g_D^*\hat{a}^\dagger - g_D\hat{a})|D\rangle\langle D|\} =\\ &\exp\{g_{AD}^*\hat{a}^\dagger - g_{AD}\hat{a}\}|A\rangle\langle D|(\hat{a}-g_D^*)\end{aligned} \tag{B6}$$

**Appendix C. The polaron transformation of the Hamiltonian**

We need to evaluate $\hat{U}\hat{H}\hat{U}^\dagger$:

$$\begin{aligned}\hat{U}\hat{H}\hat{U}^\dagger = &E_D|D\rangle\langle D| + \{E_A + \sum_n \lambda_n(\hat{b}_n^\dagger + \hat{b}_n)\}|A\rangle\langle A| + \sum_n \hbar\nu_n\hat{b}_n^\dagger\hat{b}_n\\ &+\hat{U}\{H_{DA}(|D\rangle\langle A| + |A\rangle\langle D|) + \hbar\omega\hat{a}^\dagger\hat{a} +\\ &\hbar\omega g_D(\hat{a}-\hat{a}^\dagger)|D\rangle\langle D| + \hbar\omega g_A(\hat{a}-\hat{a}^\dagger)|D\rangle\langle D| + \hbar\omega|g_D|^2|D\rangle\langle D| + \hbar\omega|g_A|^2|A\rangle\langle A| +\\ &\hbar\omega(\hat{a}-\hat{a}^\dagger)(t_{DA}|D\rangle\langle A| + t_{AD}|A\rangle\langle D|) + \hbar\omega|t_{DA}|^2 - \hbar\omega(g_D+g_A)(t_{DA}|D\rangle\langle A| + t_{AD}|A\rangle\langle D|)\}\hat{U}^\dagger\end{aligned} \tag{C1}$$

We will exercise each term in Eq. (C1) separately. Using (B4) we have:

$$\hat{U}H_{DA}(|D\rangle\langle A|+|A\rangle\langle D|)\hat{U}^\dagger =$$
$$= H_{DA}(|D\rangle\langle A|\exp\{g_{DA}^*\hat{a}^\dagger - g_{DA}\hat{a}\} + |A\rangle\langle D|\exp\{g_{AD}^*\hat{a}^\dagger - g_{AD}\hat{a}\}) \quad (C2)$$

$$\hbar\omega\hat{U}\{\hat{a}^\dagger\hat{a} + g_D(\hat{a}-\hat{a}^\dagger)|D\rangle\langle D| + g_A(\hat{a}-\hat{a}^\dagger)|D\rangle\langle D| +$$
$$|g_D|^2|D\rangle\langle D| + |g_A|^2|A\rangle\langle A|\}\hat{U}^\dagger =$$
$$= \hbar\omega\hat{U}\{(\hat{a}^\dagger + g_D)(\hat{a}+g_D^*)|D\rangle\langle D| + (\hat{a}^\dagger + g_A)(\hat{a}+g_A^*)|A\rangle\langle A|\}\hat{U}^\dagger$$
$$= \hbar\omega\{\hat{U}(\hat{a}^\dagger + g_D)|D\rangle\langle D|\hat{U}^\dagger\hat{U}(\hat{a}+g_D^*)|D\rangle\langle D|\hat{U}^\dagger \quad (C3)$$
$$+\hat{U}(\hat{a}^\dagger + g_A)|A\rangle\langle A|\hat{U}^\dagger\hat{U}(\hat{a}+g_A^*)|A\rangle\langle A|\hat{U}^\dagger\}$$
$$= \hbar\omega\hat{a}^\dagger\hat{a}$$

And using (B5) and (B6):

$$\hbar\omega\hat{U}\{(\hat{a}-\hat{a}^\dagger)(t_{DA}|D\rangle\langle A|+t_{AD}|A\rangle\langle D|) + (g_D+g_A)(t_{DA}|D\rangle\langle A|+t_{AD}|A\rangle\langle D|)\}\hat{U}^\dagger$$
$$= \hbar\omega(\hat{a}-\hat{a}^\dagger + 2g_D - g_A - g_D)t_{DA}|D\rangle\langle A|\exp\{g_{DA}^*\hat{a}^\dagger - g_{DA}\hat{a}\}$$
$$+\hbar\omega\exp\{g_{AD}^*\hat{a}^\dagger - g_{AD}\hat{a}\}(\hat{a}-\hat{a}^\dagger + 2g_D - g_A - g_D)t_{AD}|A\rangle\langle D| \quad (C4)$$
$$= \hbar\omega\{t_{DA}(\hat{a}-\hat{a}^\dagger)|D\rangle\langle A|\exp\{g_{DA}^*\hat{a}^\dagger - g_{DA}\hat{a}\} + t_{AD}\exp\{g_{AD}^*\hat{a}^\dagger - g_{AD}\hat{a}\}|A\rangle\langle D|(\hat{a}-\hat{a}^\dagger)\}$$
$$-\hbar\omega(t_{DA}g_{DA}|D\rangle\langle A|\exp\{g_{DA}^*\hat{a}^\dagger - g_{DA}\hat{a}\} - t_{AD}g_{AD}|A\rangle\langle D|\exp\{g_{AD}^*\hat{a}^\dagger - g_{AD}\hat{a}\})$$

Note that

$$(t_{DA}(\hat{a}-\hat{a}^\dagger)|D\rangle\langle A|\exp\{g_{DA}^*\hat{a}^\dagger - g_{DA}\hat{a}\})^* = t_{AD}\exp\{g_{AD}^*\hat{a}^\dagger - g_{AD}\hat{a}\}|A\rangle\langle D|(\hat{a}-\hat{a}^\dagger) \quad (C5)$$

Thus, substituting Eq. (C2) – (C4) into Eq. (C1) we have the full Hamiltonian:

$$\hat{H}' = E_D|D\rangle\langle D| + \{E_A + \sum_n \lambda_n(\hat{b}_n^\dagger + \hat{b}_n)\}|A\rangle\langle A| + \sum_n \hbar\nu_n\hat{b}_n^\dagger\hat{b}_n$$
$$+\tilde{H}_{DA}|D\rangle\langle A|\exp\{g_{DA}^*\hat{a}^\dagger - g_{DA}\hat{a}\} + \tilde{H}_{AD}|A\rangle\langle D|\exp\{g_{AD}^*\hat{a}^\dagger - g_{AD}\hat{a}\}) \quad (C6)$$
$$+\hbar\omega\hat{a}^\dagger\hat{a} +$$
$$\hbar\omega\{t_{DA}(\hat{a}-\hat{a}^\dagger)|D\rangle\langle A|\exp\{g_{DA}^*\hat{a}^\dagger - g_{DA}\hat{a}\} + t_{AD}\exp\{g_{AD}^*\hat{a}^\dagger - g_{AD}\hat{a}\}|A\rangle\langle D|(\hat{a}-\hat{a}^\dagger)\}$$

where we have introduced a new effective coupling:

$$\tilde{H}_{AD} = \tilde{H}_{DA}^* = H_{AD} + \hbar\omega t_{AD}g_{AD} \quad (C7)$$

**Appendix D. The Heisenberg equations of motion**

The Hamiltonian (19) can be re-casted onto the electronic operators $\hat{\sigma}_{ij}$:

$$\hat{H} = E_D \hat{\sigma}_{DD} + \{E_A + \sum_n \lambda_n (\hat{b}_n^\dagger + \hat{b}_n)\} \hat{\sigma}_{AA} + H_{DA}(\hat{\sigma}_{DA} + \hat{\sigma}_{AD}) + \sum_n \hbar v_n \hat{b}_n^\dagger \hat{b}_n$$
$$+ \hbar \omega \hat{a}^\dagger \hat{a} + \hbar \omega g_D (\hat{a} - \hat{a}^\dagger) \hat{\sigma}_{DD} + \hbar \omega g_A (\hat{a} - \hat{a}^\dagger) \hat{\sigma}_{AA} + \hbar \omega |g_D|^2 \hat{\sigma}_{DD} + \hbar \omega |g_A|^2 \hat{\sigma}_{AA} + \quad \text{(D1)}$$
$$\hbar \omega (\hat{a} - \hat{a}^\dagger) t_{DA} (\hat{\sigma}_{DA} - \hat{\sigma}_{AD}) + \hbar \omega |t_{DA}|^2 - \hbar \omega (g_D + g_A) t_{DA} (\hat{\sigma}_{DA} - \hat{\sigma}_{AD})$$

where

$$\hat{\sigma}_{DD} = \hat{c}_D^\dagger \hat{c}_D \quad \text{(D2a)}$$

$$\hat{\sigma}_{AA} = \hat{c}_A^\dagger \hat{c}_A \quad \text{(D2b)}$$

$$\hat{\sigma}_{AD} = \hat{\sigma}_{DA}^\dagger = \hat{c}_A^\dagger \hat{c}_D \quad \text{(D2c)}$$

and

$$\hat{\sigma}_{DD} + \hat{\sigma}_{AA} = \hat{I} \quad \text{(D2d)}$$

Here $\hat{c}_D^\dagger (\hat{c}_D)$ and $\hat{c}_A^\dagger (\hat{c}_A)$ stand for a creation(annihilation) operator of an electron on the donor and the acceptor respectively.

It is useful to employ the Pauli notation:

$$\hat{\sigma}_z = \hat{\sigma}_{AA} - \hat{\sigma}_{DD} \quad \text{(D3a)}$$

$$\hat{\sigma}_x = \hat{\sigma} + \hat{\sigma}^\dagger \quad \text{(D3b)}$$

$$\hat{\sigma}_y = i(\hat{\sigma} - \hat{\sigma}^\dagger) \quad \text{(D3c)}$$

where $\hat{\sigma} = \hat{\sigma}_{DA}$

In the present discussion the following properties of the field and electronic operators are used:

$$[\hat{a}, \hat{a}^\dagger] = 1 \quad \text{(D4a)}$$

$$[\hat{\sigma}, \hat{\sigma}^\dagger] = -\hat{\sigma}_z \quad \text{(D4b)}$$

$$[\hat{\sigma}, \hat{\sigma}_z] = 2\hat{\sigma} \quad \text{(D4c)}$$

$$[\hat{\sigma}_x, \hat{\sigma}_z] = -2i\hat{\sigma}_y \quad \text{(D4d)}$$

$$[\hat{\sigma}_y, \hat{\sigma}_z] = 2i\hat{\sigma}_x \quad \text{(D4e)}$$

$$[\hat{\sigma}_x, \hat{\sigma}_y] = 2i\hat{\sigma}_z \quad \text{(D4f)}$$

Using Eqs (D4), we obtain the Heisenberg equations of motion for the EM field, electronic and bath operators:

$$\frac{d\hat{a}}{dt} = \frac{1}{i\hbar}[\hat{a}, \hat{H}] = -i\omega\hat{a} + i\omega t_{DA}(\hat{\sigma}_{DA} - \hat{\sigma}_{AD}) +$$
$$i\omega g_D \hat{\sigma}_{DD} + i\omega g_A \hat{\sigma}_{AA} = -i\omega\hat{a} + \omega t_{DA}\hat{\sigma}_y + i\omega(g_D + g_A + g_{DA}\hat{\sigma}_z)/2 \quad \text{(D5a)}$$
$$= -i\omega\{\hat{a} - (g_D + g_A + g_{DA}\hat{\sigma}_z)/2\} + \omega t_{DA}\hat{\sigma}_y$$

$$\frac{d\hat{\sigma}_z}{dt} = \frac{1}{i\hbar}[\hat{\sigma}_z, \hat{H}] = [\hat{\sigma}_z, -iH_{DA}/\hbar\hat{\sigma}_x - \omega(\hat{a} - \hat{a}^\dagger)t_{DA}\hat{\sigma}_y + \omega(g_D + g_A)t_{DA}\hat{\sigma}_y] =$$
$$2H_{DA}/\hbar\hat{\sigma}_y + 2i\omega(\hat{a} - \hat{a}^\dagger - g_D - g_A)t_{DA}\hat{\sigma}_x \quad \text{(D5b)}$$

$$\frac{d\hat{\sigma}_x}{dt} = \frac{1}{i\hbar}[\hat{\sigma}_x, \hat{H}] = \frac{1}{2i\hbar}(\{E_A + \sum_n \lambda_n(\hat{b}_n^\dagger + \hat{b}_n) - E_D + \hbar\omega(g_A - g_D)(\hat{a} - \hat{a}^\dagger - g_A - g_D)\}$$
$$\times[\hat{\sigma}_x, \hat{\sigma}_z] - 2i\hbar\omega(\hat{a} - \hat{a}^\dagger - g_A - g_D)t_{DA}[\hat{\sigma}_x, \hat{\sigma}_y]) =$$
$$= -\frac{1}{\hbar}\hat{\sigma}_y\{E_A + \sum_n \lambda_n(\hat{b}_n^\dagger + \hat{b}_n) - E_D + \hbar\omega(g_A - g_D)(\hat{a} - \hat{a}^\dagger - g_A - g_D)\} - \quad \text{(D5c)}$$
$$2i\omega(\hat{a} - \hat{a}^\dagger - g_A - g_D)t_{DA}\hat{\sigma}_z$$

$$\frac{d\hat{\sigma}_y}{dt} = \frac{1}{i\hbar}[\hat{\sigma}_y, \hat{H}] = \frac{1}{2i\hbar}(\{E_A + \sum_n \lambda_n(\hat{b}_n^\dagger + \hat{b}_n) - E_D + \hbar\omega(g_A - g_D)(\hat{a} - \hat{a}^\dagger - g_A - g_D)\}$$
$$\times[\hat{\sigma}_y, \hat{\sigma}_z] + 2H_{DA}[\hat{\sigma}_y, \hat{\sigma}_x]) = \quad \text{(D5d)}$$
$$\frac{1}{\hbar}\hat{\sigma}_x\{E_A + \sum_n \lambda_n(\hat{b}_n^\dagger + \hat{b}_n) - E_D + \hbar\omega(g_A - g_D)(\hat{a} - \hat{a}^\dagger - g_A - g_D)\} - \frac{2H_{DA}}{\hbar}\hat{\sigma}_z$$

$$\frac{d\hat{b}_n}{dt} = -i\nu_n\hat{b}_n - i\lambda_n/\hbar\hat{\sigma}_{AA} = -i\nu_n\hat{b}_n - i\frac{\lambda_n}{2\hbar} - i\frac{\lambda_n}{2\hbar}\hat{\sigma}_z \quad \text{(D5e)}$$

We redefine the field operator $\hat{a}$ as follows:

$$\hat{a} - (g_D + g_A + g_{DA}\hat{\sigma}_z)/2 \to \hat{a} \quad \text{(D6)}$$

which corresponds to the polaron transformation (except the term $g_{DA}\hat{\sigma}_z$) discussed in Appendix C and implies that the photon "follows" the electron position. Note that such substitution is possible since $\hat{\sigma}_z$ is self-adjoint and $[\hat{a}, \hat{\sigma}_z] = 0$ otherwise (D6) would not be a bosonic operator. With (D6) Eqs. (D5) are re-written as follows:

$$\frac{d\hat{a}}{dt} = -i\omega\hat{a} + \omega t_{DA}\hat{\sigma}_y - \frac{1}{2}g_{DA}\frac{d\hat{\sigma}_z}{dt} \quad \text{(D7a)}$$

$$\frac{d\hat{\sigma}_z}{dt} = \frac{1}{i\hbar}[\hat{\sigma}_z, \hat{H}] = 2H_{DA}/\hbar\hat{\sigma}_y + 2i\omega(\hat{a} - \hat{a}^\dagger + g_{DA}\hat{\sigma}_z)t_{DA}\hat{\sigma}_x \quad \text{(D7b)}$$

$$\frac{d\hat{\sigma}_x}{dt} = -\frac{1}{\hbar}\hat{\sigma}_y\{E_A + \sum_n \lambda_n(\hat{b}_n^\dagger + \hat{b}_n) - \sum_n \frac{\lambda_n^2}{\hbar\nu_n} - E_D - \hbar\omega g_{DA}(\hat{a} - \hat{a}^\dagger + g_{DA}\hat{\sigma}_z)\} - \quad \text{(D7c)}$$
$$2i\omega(\hat{a} - \hat{a}^\dagger + g_{DA}\hat{\sigma}_z)t_{DA}\hat{\sigma}_z$$

$$\frac{d\hat{\sigma}_y}{dt} = \frac{1}{\hbar}\hat{\sigma}_x\{E_A + \sum_n \lambda_n(\hat{b}_n^\dagger + \hat{b}_n) - \sum_n \frac{\lambda_n^2}{\hbar\nu_n} - E_D - \hbar\omega g_{DA}(\hat{a} - \hat{a}^\dagger + g_{DA}\hat{\sigma}_z)\} + \quad \text{(D7d)}$$
$$-\frac{2H_{DA}}{\hbar}\hat{\sigma}_z$$

$$\frac{d\hat{b}_n}{dt} = -i\nu_n\hat{b}_n - i\frac{\lambda_n}{2\hbar}\hat{\sigma}_z \quad \text{(D7e)}$$

where the substitution $\hat{b}_n \to \hat{b}_n + \frac{\lambda_n}{2\hbar\nu_n}$ has also been made. From Eqs. (D7) it is clear that the system dynamics depends only on the relative distance between the donor and the acceptor but not on their positions. Also, a reorganization energy term $\sum_n \frac{\lambda_n^2}{\hbar\nu_n}$ appears in the Eqs.(D7c-D7d). In other words, the polaron transformation can be seen as a change of variables in the Heisenberg equations of motion for the bosonic mode operators.

Equation (D7a) can be easily solved as follows:

$$\hat{a}(t) = \hat{a}\exp(-i\omega t) + \omega\int_0^t \exp\{-i\omega(t-t')\}t_{DA}\hat{\sigma}_y dt' - \frac{1}{2}\int_0^t \exp\{-i\omega(t-t')\}g_{DA}\frac{d\hat{\sigma}_z}{dt'}dt' \quad \text{(D8)}$$

Consider the last integral. Suppose $\frac{d\hat{\sigma}_z}{dt'}$ is varying slowly on a time scale $T$ (tunneling time) such as that $T\omega \gg 1$. Because of the fast oscillating exponent $\exp[-i\omega(t-t')]$, the integral is zero. Thus, we arrive at the standard expression[87] for the operator of the EM field interacting with a two-level system:

$$\hat{a}(T) = \hat{a}\exp(-i\omega T) + \omega\int_0^T \exp\{-i\omega(t-t')\}t_{DA}\hat{\sigma}_y dt' \quad \text{(D9)}$$

This result is expected: the change of variables given by (D6) automatically provides such form of equations of motion where the photon is always adjusted to the electron position if the electron is slow.

On the other hand, if $\frac{d\hat{\sigma}_z}{dt'}$ varies rapidly during the tunneling and $T\omega \ll 1$ then $\exp\{-i\omega(t-t')\} \sim 1$ and Eq.(D9) takes the following form:

$$\hat{a}(T) = \{\hat{a} - g_{DA}\{\hat{\sigma}_z(T) - \hat{\sigma}_z(0)\}/2\} + \omega \int_0^T t_{DA}\hat{\sigma}_y dt' \quad (D10)$$
$$\approx \hat{a} - g_{DA}\{\hat{\sigma}_z(T) - \hat{\sigma}_z(0)\}/2\}$$

Here it has been assumed that the last term in Eq. (D10) can be neglected. Indeed, for the electron transfer problem dipole coupling between diabatic states $t_{DA}$ is small and $\omega T \ll 1$ which leads to the second line in (D10).

From Eq. (D10) it is clear that the photon mode is shifted. Thus, an eigenvector of the mode after the tunneling corresponds to this displacement:

$$\hbar\omega Tr_e\{\rho_e(0)\hat{a}^\dagger(T)\hat{a}(T)\}\psi_n(T) =$$
$$\hbar\omega\{\hat{a}^\dagger - Tr_e(\rho_e(0)g_{AD}\{\hat{\sigma}_z(T) - \hat{\sigma}_z(0)\}/2)\}\{\hat{a} - Tr_e(\rho_e(0)g_A\{\hat{\sigma}_z(T) - \hat{\sigma}_z(0)\}/2)\} \quad (D11)$$
$$= \hbar\omega\{\hat{a}^\dagger - g_{AD}\}\{\hat{a} - g_{DA}\}\psi_n(T) = n\hbar\omega\psi_n(T)$$

where the trace is over the electron states and $\rho_e(0)$ is an initial electron density and the operator $\hat{\sigma}_z(0)$ relates to the electron residing on the donor whereas $\hat{\sigma}_z(T)$ corresponds to the electron on the acceptor. Note that $\hbar\omega Tr_e\{\rho_e(0)\hat{a}^\dagger(t)\hat{a}(t)\}$ is an effective time-dependent Hamiltonian in the reduced space of the mode.

From Eq. (D11) it follows that

$$\psi_n(T) = \exp(\hat{a}^\dagger g_{DA} - \hat{a}g_{AD})\psi_n(0) \quad (D12)$$

where $\exp(\hat{a}^\dagger g_{DA} - \hat{a}g_{AD})$ is a shift operator. In the Heisenberg picture transition probabilities are defined by overlaps between the eigenstates which evolve over time and the time-independent wavefunction. Thus, the overlap between the eigenstates after and before the tunneling gives an electromagnetic Frank-Condon factor:

$$\langle\psi_n(T)|\psi_{n'}(0)\rangle = F_{nn'}(g_{AD}) \quad (D13)$$

which agrees with the result (21). The term $\omega\int_0^T t_{DA}\hat{\sigma}_y dt'$ omitted in (D10) can be then reintroduced as the standard dipole coupling modified by electromagnetic FC factors.

[67] In principle, all electrostatic interactions are mediated by the electromagnetic field, so $V_{coul}$ in Eq. (4) may be thought of as due to all EM modes except the lowest cavity-supported mode that is considered explicitly.